# Analytic Approximation of Free-Space Path Loss for Implanted Antennas

MINGXIANG GAO[1], SUJITH RAMAN[1,3] (Senior Member, IEEE), ZVONIMIR SIPUS[2] (Senior Member, IEEE), AND ANJA K. SKRIVERVIK[1]

[1] Microwaves and Antennas Group, Ecole Polytechnique Federale de Lausanne, Lausanne, Switzerland
[2] Faculty of Electrical Engineering and Computing, University of Zagreb, Zagreb, Croatia
[3] Radio Systems Group, University of Twente, Enschede, Netherlands

CORRESPONDING AUTHOR: ANJA K. SKRIVERVIK (e-mail: anja.skrivervik@epfl.ch).

This work was supported in part by the Croatian Science Foundation (HRZZ).

**ABSTRACT** Implantable wireless bioelectronic devices enable communication and/or power transfer through RF wireless connections with external nodes. These devices encounter notable design challenges due to the lossy nature of the host body, which significantly diminishes the radiation efficiency of the implanted antenna and tightens the wireless link budget. Prior research has yielded closed-form approximate expressions for estimating losses occurring within the lossy host body, known as the in-body path loss. To assess the total path loss between the implanted transmitter and external receiver, this paper focuses on the free-space path loss of the implanted antenna, from the body–air interface to the external node. This is not trivial, as in addition to the inherent radial spreading of spherical electromagnetic waves common to all antennas, implanted antennas confront additional losses arising from electromagnetic scattering at the interface between the host body and air. Employing analytical modeling, we propose closed-form approximate expressions for estimating this free-space path loss. The approximation is formulated as a function of the free-space distance, the curvature radius of the body–air interface, the depth of the implanted antenna, and the permittivity of the lossy medium. This proposed method undergoes thorough validation through numerical calculations, simulations, and measurements for different implanted antenna scenarios. This study contributes to a comprehensive understanding of the path loss in implanted antennas and provides a reliable analytical framework for their efficient design and performance evaluation.

## I. INTRODUCTION

Implantable bioelectronics have undergone rapid development since the inception of the first artificial cardiac pacemaker was implanted in 1958 [1], allowing for direct interfacing with biological tissues and organs. Today, implantable bioelectronics are widely used in monitoring and treating a range of medical conditions, including heart disease, neurological disorders, hearing impairment, etc [2], [3], [4]. These devices are minimally invasive, customizable to individual patients, and offer the potential for prolonged monitoring and treatment. Given these characteristics, the long-term stability and sustainability of implantable bioelectronics require efficient wireless communication and power supply [6], [7], [3]. Consequently, understanding the loss mechanisms and evaluating the path loss of implanted antennas becomes critical for achieving optimal wireless links.

Despite the widespread adoption of wireless systems due to advancements in radio frequency (RF) technology, establishing wireless connections for implantable devices remains a formidable challenge. The primary obstacle lies in the significant electromagnetic absorption and losses caused by highly lossy biological tissues [8], [9], [10]. In the context of implanted antennas, electromagnetic radiation cannot be separated from the surrounding medium and is strongly affected by near-field coupling, propagation attenuation, and interaction with the body interfaces [11], [12]. Additionally, the miniaturized dimensions of implantable bioelectronics impose further constraints on the radiation performance of implanted antennas [13].

For a transmitter and a receiver in free space, the Friis transmission equation is commonly used to estimate wireless transmission efficiency, i.e., the free-space path loss formula, combined with the gains of the antennas [14], [15]. This approach quantitatively assesses losses or gains from various factors, providing crucial guidance in the initial phases of wireless system design. However, when dealing with an implantable wireless system that includes an implanted antenna and an external antenna, quick path loss estimates must account for the effects of the lossy host body. Ideally, with the gain of the implanted antenna available, the Friis formula could still be employed for transmission efficiency estimation. Unfortunately, evaluating antenna gain remains challenging during the early stages of implanted antenna design, which is indeed a pivotal parameter for optimization. A pragmatic approach is to consider the implanted antenna itself (excluding the host body) as the source of interest and analyze the EM radiation in lossy media and free space. Therefore, the primary research objective is to gain theoretical insights into the path loss between the implanted antenna and an external node. As shown in Fig. 1, the path loss for a typical implanted antenna comprises in-body path loss and free-space path loss.

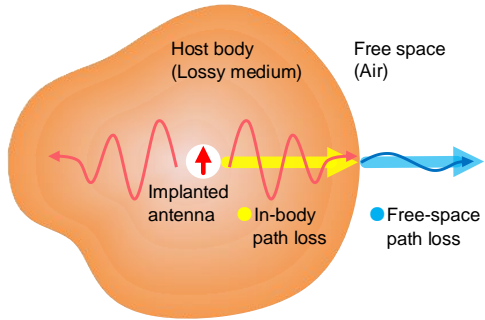

**FIGURE 1.** Sketch of an antenna implanted in a lossy medium.

To assess the path loss of an implanted antenna under a specific implantation condition, analytical modeling is preferred to give physical insights into the loss mechanisms within different regions or media. Indeed, analytical modeling, a classic approach to antenna analysis and design, offers a pathway to understanding the physical mechanisms and performance characteristics of antennas [16], [17], [18]. Therefore, this approach is very useful in providing crucial benchmarks and optimization strategies for antenna design. As for implanted antennas, analytical analysis of simplified body models offers a macroscopic perspective to investigate the electromagnetic (EM) radiation [19], [20] and wireless power transmission [21] characteristics of implants, with a special focus on near-field losses for different encapsulations [22], [23].

Studies have been carried out first on the in-body EM wave propagation and in-body path loss of implanted antennas through an analytical modeling approach [24], [25]. In-body path loss assesses losses along the shortest wireless link, i.e., from the implanted antenna to the body–air interface [26], [27], [28]. This path loss contributes to a major attenuation of EM radiation from implants compared to antennas operating in free space. In-body path loss can be expressed as the product of three main contributions: losses due to dissipation in the reactive near field, propagating field absorption, and reflections at the body–air interface [24]. Specifically, the derived approximate expressions for in-body path loss are fully derived to estimate the maximum power density reaching free space from a transmitting antenna implanted in a large-scale host body [25]. However, there is a lack of insight into the free-space path loss. The conventional free-space path loss formula cannot be directly applied to the wave propagation estimation from the host body interface to free space. This is due to the fact that complex EM scattering at the body interface makes the entire interface an equivalent aperture radiating into free space. As a result, the host body not only directly causes significant EM absorption losses, but also requires attention to the beam deformation of the implanted antenna due to its high permittivity property.

This paper aims to investigate an analytical model of an implanted antenna that provides closed-form approximate expressions for the total path loss from an implant to an external node in free space. One important purpose of this model is to provide a realistic estimation of the best possible achievable wireless link between an implant and an external node prior to any design, and thus a benchmark to be used during the actual design process. In particular, characteristics of the implanted antenna and its host body, including the curvature radius of the body–air interface, permittivity of the host body, and implantation depth, are additionally taken into account in the approximate calculation of the free-space path loss. The proposed method is validated with multiple cases of implanted antennas.

## II. ANALYTICAL MODELING AND DERIVATION

Path loss is a key performance indicator for evaluating the losses of implantable antennas. Unlike wireless links between devices operating in free space, the path loss from the implantable device to the external node includes the in-body path loss and the free-space path loss. To maximize link efficiency, the path loss analyzed here is from the implant through the closest body interface to free space. In-body path loss, evaluating the path loss over the shortest wireless link from the implant to the body–air interface, represents the inevitable losses absorbed by the host body, and an analytic approximation of it can be found in [25]. It is related to the upper limit of the wireless transmission efficiency from the implant to the body–air interface.

For the external node placed at a certain distance from the host body, the free-space path loss from the body–air interface to the external node is also affected by the host body. According to the equivalence theorem, the EM waves radiated by the host body can be regarded as radiated by the equivalent surface currents on the body–air interface, which are excited by the implanted antenna. For small-scale host bodies, the dielectric resonator effect can disrupt the current distribution on the body–air interface, resulting in the deformation of the radiation pattern [29]. For large-scale host bodies, analytical modeling is proposed to estimate the free-space path loss as a function of the characteristics of the host body.

To this aim, we consider a canonical model: an elementary EM source placed in an electrically large spherical body model, as shown in Fig. 2. The body model represents a large-scale host body, and its radius $r_{body}$ is used to model the curvature radius of the host body interface close to the implant. When the body interface approaches a plane, $r_{body}$ becomes very large, and the spherical body model starts to approach a planar body model. To simulate the biological tissue, the complex permittivity $\varepsilon_r$ evaluated by the four-region Cole–Cole model [30] is used as the lossy medium in the modeling. Moreover, the body phantom is set to be homogeneous to facilitate initial analysis.

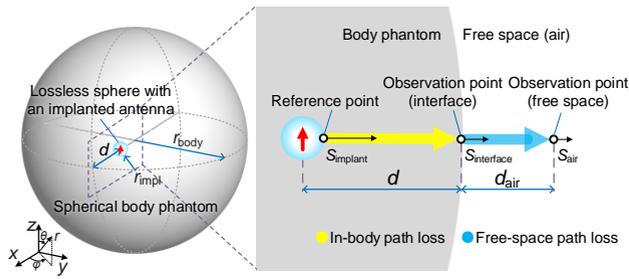

**FIGURE 2.** View of a spherical body model with an elementary dipole source implanted at a depth of *d*.

The implanted antenna is modeled as an elementary electric or magnetic dipole surrounded by a small air sphere of radius $r_{\text{impl}}$, which roughly represents the dimension of the implant encapsulation and is typically shorter than the wavelength in the body media $\lambda_{\text{body}}$. The implantation depth $d$ represents the distance from the implant to the closest body–air interface. According to the analysis of dipole radiation in a lossy medium [31], maximum gain occurs when the source is oriented parallel to the planar body interface. In the case of the large-scale spherical body model analyzed here, this means that the dipole source is oriented in the $z$-direction, i.e., in parallel to the closest body–air interface. In analyzing the implanted antenna, it is assumed that the dipole is well-matched and has no conductor losses or dielectric losses due to the encapsulation.

First, we review the analytic approximations to assess the in-body path loss of implanted antennas, i.e., the path loss from the implant to the observation point at the closest body–air interface (see Fig. 2):

$$\text{Re}\{S_{\text{interface}}\} = \text{Re}\{S_{\text{implant}}\} \cdot \frac{r_{\text{impl}}^2}{d^2} \cdot e_{\text{total}}$$
$$= \text{Re}\{S_{\text{implant}}\} \cdot \frac{r_{\text{impl}}^2}{d^2} \cdot e_{\text{near field}} \cdot e_{\text{propagation}} \cdot e_{\text{reflection}}, \quad (1)$$

where $\text{Re}\{S_{\text{implant}}\}$ and $\text{Re}\{S_{\text{interface}}\}$ represent the power density of EM waves entering the body (i.e., the reference point on the implant encapsulation) and reaching the body–air interface (i.e., the observation point on the interface), respectively, $r_{\text{impl}}^2/d^2$ accounts for the effect of the radial spreading of spherical EM waves, and $e_{\text{total}}$ represents the total path loss after excluding the radial spreading of spherical EM waves. From the implant to the body–air interface, $e_{\text{total}}$ can be decomposed into three different loss contributions. $e_{\text{near field}}$ accounts the losses caused by the coupling of the reactive near field of the implanted antenna with the lossy medium, $e_{\text{propagation}}$ represents the losses due to the propagating field absorption in the lossy medium, and $e_{\text{reflection}}$ accounts for the reflection losses at the body–air interface. Specific closed-form expressions for these loss terms can be found in [24] and [25].

In this work, we focus on the free-space path loss from the body–air interface to the observation point in free space. To assess maximum link efficiency, this external observation point is along the direction from the implant to the closest body–air interface, as shown in Fig. 2. The distance from the body–air interface to the observation point in free space is denoted as $d_{\text{air}}$. From the perspective of wave propagation, the spherical waves radiated by the implanted antenna will undergo significant refraction when reaching the body–air interface. In particular, the permittivity of most biological tissues can be relatively high due to their high water content. As a result, refraction occurring at the body–air interface causes further spreading of the propagating spherical waves reaching free space. In addition to propagating fields, for cases with shallow implant depths ($d < \lambda_{\text{body}}$), the near field reaching the body interface can partially enhance the EM waves in the "near field region" of the body–air interface, but cannot propagate to the far field. For the path loss from the implant to the observation point in free space, (1) is rewritten as

$$\text{Re}\{S_{\text{air}}\} = \text{Re}\{S_{\text{implant}}\} \frac{r_{\text{impl}}^2}{(d + d_{\text{air}})^2} \cdot e_{\text{total}}$$
$$= \text{Re}\{S_{\text{implant}}\} \frac{r_{\text{impl}}^2}{(d + d_{\text{air}})^2} \cdot e_{\text{near field}} \cdot e_{\text{propagation}} \cdot e_{\text{reflection}} \cdot e_{\text{refraction}}, \quad (2)$$

where $\text{Re}\{S_{\text{air}}\}$ represents the power density of EM waves reaching the observation point in free space, and $e_{\text{refraction}}$ represents the losses due to wave scattering at the body–air interface. Here, we only consider the path loss of the propagating wave since the external receiver is usually located in the "far-field region" of the body–air interface.

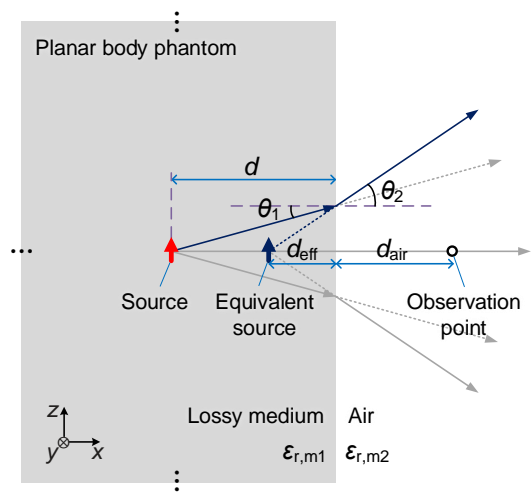

**FIGURE 3.** Schematic illustration of the refraction process at a planar body–air interface, where spherical EM waves are radiated by an implanted antenna.

Next, we derive the efficiency term $e_{\text{refraction}}$ by analyzing the refraction process of EM waves at the body–air interface.

In the case of a planar body–air interface, the refraction process of spherical waves excited from the implanted source is illustrated in Fig. 3. For EM waves with an angle of incident angle $\theta_1$, waves propagating into free space have an angle of refraction $\theta_2$.

According to Snell's law for refraction [32], the angle of incidence $\theta_1$ and angle of refraction $\theta_2$ are linked by

$$\frac{\sin\theta_1}{\sin\theta_2} = \frac{n_2}{n_1} = \frac{v_{p,m1}}{v_{p,m2}} = \frac{k'_{m2}}{k'_{m1}}, \quad (3)$$

where $n$ represents the refractive indices, $v_p$ is the phase velocity in the considered medium, and $k'$ is the real part of the complex wavenumber of the considered medium.

For the observation point in free space (as indicated in Fig. 3), it meets the condition of $\theta_1 \to 0$ and $\theta_2 \to 0$. Thus, the ratio between $\theta_1$ and $\theta_2$ can be further written as

$$\frac{\sin\theta_1}{\sin\theta_2} \approx \frac{\tan\theta_1}{\tan\theta_2} \approx \frac{\theta_1}{\theta_2}. \quad (4)$$

Under the same conditions, an equivalent source can be obtained by reverse extension of the EM waves propagating in free space. As shown in Fig. 3, the corresponding equivalent implantation depth is denoted as $d_{\text{eff}}$. According to the geometric relationship,

$$d \tan\theta_1 = d_{\text{eff}} \tan\theta_2. \quad (5)$$

Thus, we can get

$$d_{\text{eff}} = \frac{\tan\theta_1}{\tan\theta_2} d \approx \frac{\sin\theta_1}{\sin\theta_2} d \approx \frac{k'_{m2}}{k'_{m1}} d. \quad (6)$$

For the observation point in free space, i.e., $d_{\text{air}} \geq 0$, the spreading of spherical EM waves (i.e., the outward propagating spherical wave attenuates by the square of the spreading distance) needs to be corrected in order to include the losses due to wave scattering at the body–air interface. By moving the original source at depth $d$ to the equivalent source at depth $d_{\text{eff}}$ and transforming the spreading distance from $d + d_{\text{air}}$ to $d_{\text{eff}} + d_{\text{air}}$, $e_{\text{refraction}}$ can be calculated as

$$e_{\text{refraction}} = \frac{(d + d_{\text{air}})^2 d_{\text{eff}}^2}{d^2 (d_{\text{eff}} + d_{\text{air}})^2}. \quad (7)$$

The above expression only takes into account the additional losses caused by refraction at the body–air interface. Thus, as shown in Fig. 3, the radial propagation spreading of power density in free space is recalculated by taking the value of power density at the body–air interface, but using the effective origin (i.e., the position of the equivalent source) to calculate the spreading in free space. The radial propagation spread of the power density has indeed been expressed by the term $r_{\text{impl}}^2 / (d + d_{\text{air}})^2$ in (2).

For observation points located in the far-field region of the body–air interface, i.e., $d_{\text{air}} \to \infty$, $e_{\text{refraction}}$ approaches a constant

$$e_{\text{refraction}} = \frac{d_{\text{eff}}^2}{d^2} \approx \left(\frac{k'_{m2}}{k'_{m1}}\right)^2. \quad (8)$$

In calculating the wavenumber for the lossy media m1 [33], an approximation can be applied for the case when $\varepsilon'_{r,m1} > 3\varepsilon''_{r,m1}$, given as

$$k'_{m1} = \omega\sqrt{\mu_0\varepsilon_0}\,\text{Re}\left(\sqrt{\varepsilon'_{r,m1} - i\varepsilon''_{r,m1}}\right) \approx \omega\sqrt{\mu_0\varepsilon_0\varepsilon'_{r,m1}}. \quad (9)$$

Therefore, for lossy body medium that satisfies $\varepsilon'_{r,m1} > 3\varepsilon''_{r,m1}$ and the external lossless medium (air) with $\varepsilon_{r,m2} = 1$, $e_{\text{interface}}$ can be further simplified as

$$e_{\text{refraction}} \approx \frac{1}{\varepsilon'_{r,m1}}. \quad (10)$$

In the case of the spherical body–air interface, the refraction process of spherical waves excited from the implanted source is illustrated in Fig. 4. Although the ratio of $\sin\theta_1$ and $\sin\theta_2$ still satisfies (3), the curvature of the body–air interface needs to be taken into account, which ultimately changes the location of the equivalent source.

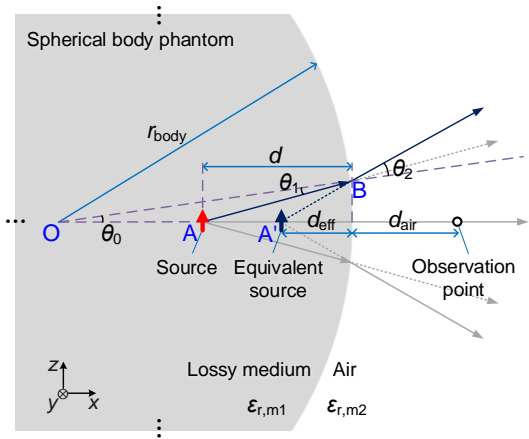

**FIGURE 4.** Schematic illustration of the refraction process at a spherical body–air interface, where spherical EM waves are radiated by an implanted antenna.

Based on the geometry shown in Fig. 4, using the sine theorem, the following system of equations can be established as

$$\begin{cases} \dfrac{r_{\text{body}} - d}{r_{\text{body}}} = \dfrac{\sin\theta_1}{\sin(\theta_0 + \theta_1)} \\[2mm] \dfrac{r_{\text{body}} - d_{\text{eff}}}{r_{\text{body}}} = \dfrac{\sin\theta_2}{\sin(\theta_0 + \theta_2)} \\[2mm] \dfrac{\sin\theta_1}{\sin\theta_2} \approx \dfrac{k'_{m2}}{k'_{m1}} \\[2mm] \theta_1 \to 0, \theta_2 \to 0 \end{cases} \quad (11)$$

where the first two equations are derived by applying the sine theorem to triangles (O, A, B) and (O, A', B) in Fig. 4. The resulting expression of the equivalent implantation depth is

$$d_{\text{eff}} \approx \frac{\theta_0 + \theta_1}{\theta_0 + \theta_2} d \approx \frac{r_{\text{body}} d}{\frac{k'_{m1}}{k'_{m2}}(r_{\text{body}} - d) + d}. \quad (12)$$

Similarly, $e_{\text{refraction}}$ can be expressed as (7) by transforming the spreading distance from $d + d_{\text{air}}$ to $d_{\text{eff}} + d_{\text{air}}$.

For observation points in the far-field region of the body–air interface, i.e., $d_{\text{air}} \to \infty$, $e_{\text{refraction}}$ approaches to a constant value

$$e_{\text{refraction}} = \frac{d_{\text{eff}}^2}{d^2} \approx \frac{r_{\text{body}}^2}{\left[\frac{k'_{m1}}{k'_{m2}}(r_{\text{body}} - d) + d\right]^2}. \quad (13)$$

## III. NUMERICAL VALIDATIONS

In this section, three examples of implantable antennas are investigated through numerical calculations and simulations. The derived approximate expressions are verified in estimating the path loss in free space, i.e., from the body–air interface to the far-field region in free space.

### A. IMPLANTED ANTENNAS IN PLANAR BODY MODELS

We first consider a canonical model to validate the usefulness of the proposed method in planar body models: an elementary EM source placed in a planar body phantom, as shown in Fig. 5.

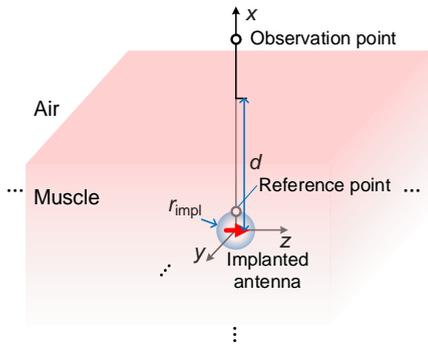

**FIGURE 5.** View of a planar body model with an elementary dipole source implanted at a depth of *d*.

In this model, the host body is modeled as a planar body phantom, i.e., a semi-infinite lossy medium with a planar interface. To simulate the biological tissue, the permittivity of the lossy medium is set equal to that of muscle [30]. Similar to the modeling in Fig. 2, the implanted miniature antenna is modeled as an elementary electric dipole surrounded by a small air sphere of radius $r_{\text{impl}} = 1$ mm. The dipole is oriented in the *z*-direction, i.e., parallel to the body–air interface for maximum gain in the +*x*-direction. The implantation depth *d*, representing the distance from the implant to the nearest body–air interface, is set to 3, 5, and 7 cm in the analysis, respectively. For biomedical purposes, the operating frequency of the antenna is 2.45 GHz, which is within the industrial, scientific, and medical (ISM) band of 2.4 to 2.5 GHz.

In the numerical calculation, Green's functions method for multilayered planar media is used to accurately calculate the EM fields and power densities from the body phantom to free space [34], [35], [36]. In order to analyze the path loss of the implanted antenna along the *x*-axis direction, we directly calculate the power density of the excited EM waves at the reference point and the observation point, i.e., $\text{Re}\{S_{\text{implant}}\}$ and $\text{Re}\{S_{\text{interface}}\}$. By translating the observation point along the *x*-axis, the total path loss excluding radial spreading $e_{\text{total}}$ can be calculated as

$$e_{\text{total}} = \frac{\text{Re}\{S(x)\}}{\text{Re}\{S_{\text{implant}}\}} \frac{x^2}{r_{\text{impl}}^2}, \quad (14)$$

which is a function of the coordinate distance *x* valid both in the body and in free space. In Fig. 6, the numerical results are shown as black solid lines.

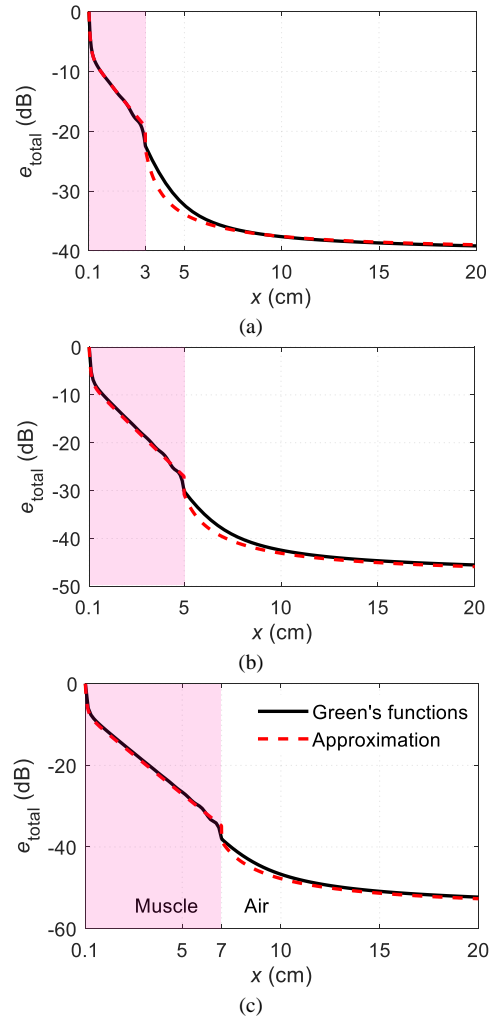

**FIGURE 6.** The total path loss $e_{\text{total}}$ as a function of the coordinate distance *x*, where the implantation depth *d* is (a) 3 cm, (b) 5 cm, and (c) 7 cm, respectively.

Another approach to calculate $e_{total}$ is to use the proposed analytical approximation of different loss contributions, as illustrated in (1) and (2). Specifically, the total path loss excluding radial spreading $e_{total}$ can be expressed by analytical expressions as

$$e_{total} = \begin{cases} e_{near\,field} \cdot e_{propagation}, & r_{impl} \leq x < d \\ e_{near\,field} \cdot e_{propagation} \cdot e_{reflection} \cdot e_{refraction}, & x \geq d \end{cases}. \quad (15)$$

Note that when calculating the power density within the body phantom, i.e, $r_{impl} \leq x < d$, $d$ in the expressions of $e_{near\,field}$ and $e_{propagation}$ needs to be replaced with $x$.

The approximate results of $e_{total}$ are demonstrated as red dashed lines in Fig. 6. As shown in Fig. 6, for different implantation depths, the path loss calculated by Green's functions is generally in good agreement with the approximate results, which verifies the proposed approximation method. In particular, for path losses in free space (air), approximate results almost overlap with numerical calculations in the far-field region of the body–air interface. For the case where the implantation depth $d = 3$ cm, the near field reaches the body–air interface after experiencing strong coupling loss. As a result, the power density is enhanced in the near-field region of the interface, and the deviation between the approximate and Green's function results can be up to 3 dB. As the observation point moves to the far-field region, this deviation decreases, ultimately achieving an effective estimation of the path loss.

### B. AN IMPLANTED ANTENNA IN A PLANAR BODY MODEL WITH A MATCHING LAYER

Based on the planar body model, we further analyze a practical multi-layered planar model, that is, a planar model with a matching layer for improving link efficiency.

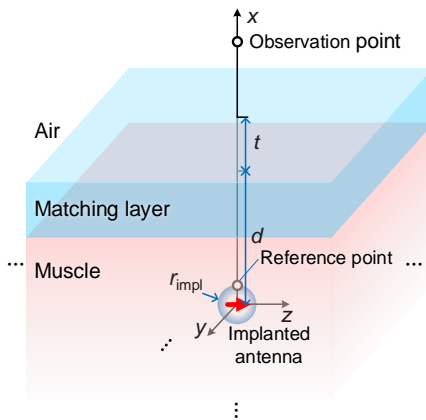

**FIGURE 7.** View of a planar body model with a matching layer in which an elementary dipole source is implanted at a depth of $d$.

As shown in Fig. 7, the host body is also modeled as a planar body phantom, and the permittivity of the lossy medium is set equal to that of muscle to simulate biological tissue. In order to reduce the severe reflection loss occurring at the body–air interface, a practical measure is to add a matching layer with a quarter-wavelength thickness at the body interface. According to the principle of a quarter-wavelength transformer, the matching layer should be made of a lossless medium, and its ideal permittivity is the square root of the host body's permittivity [37], [38].

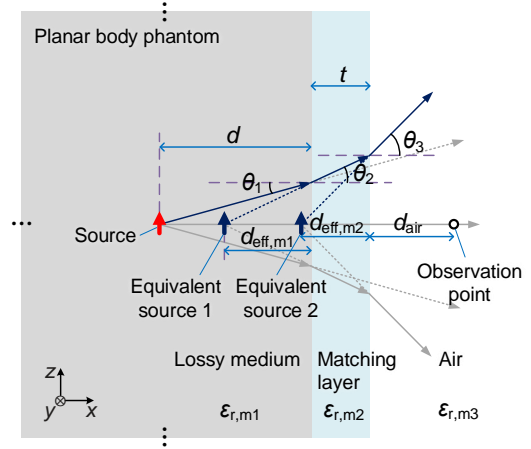

**FIGURE 8.** Schematic illustration of the refraction process in a planar body model with a matching layer, where spherical EM waves are radiated by an implanted antenna.

In the case of a 2-layer planar body model, the refraction process of spherical waves excited from the implanted source is illustrated in Fig. 8. Due to the existence of two planar interfaces, the refraction process needs to be divided into two parts, contributing to equivalent source 1 and equivalent source 2, respectively. Using the derivation method given in Section II, the corresponding equivalent implantation depths are

$$d_{eff,m1} \approx \frac{k'_{m2}}{k'_{m1}} d \quad \text{and} \quad d_{eff,m2} \approx \frac{k'_{m3}}{k'_{m1}} (d_{eff,m1} + t). \quad (16)$$

Similar to (15), the total path loss excluding radial spreading $e_{total}$ can be further derived as

$$e_{total} = \begin{cases} e_{near\,field} \cdot e_{propagation}, & r_{impl} \leq x < d \\ e_{near\,field} \cdot e_{propagation} \cdot e_{reflection} \cdot e_{refraction}^{m2}, & d \leq x < d+t \\ e_{near\,field} \cdot e_{propagation} \cdot e_{reflection} \cdot e_{refraction}^{m3}, & x \geq d+t \end{cases},$$

(17)

where

$$e_{refraction}^{m2} \approx \left(\frac{k'_{m2}}{k'_{m1}}\right)^2 \frac{x^2}{(d_{eff,m1} + x - d)^2}, \quad (18)$$

$$e_{refraction}^{m3} \approx \left(\frac{k'_{m3}}{k'_{m1}}\right)^2 \frac{x^2}{(d_{eff,m2} + d_{air})^2}. \quad (19)$$

In evaluating the reflection loss $e_{reflection}$ at the body–matching layer interface, the input impedance of the matching layer loaded with air (denoted as $Z_{in}$) needs to be considered. Thus, we have

$$e_{\text{reflection}} \approx \frac{\text{Re}\left[\left|2Z_{\text{in}}/(Z_{\text{in}}+Z_{\text{m1}})\right|^2 / Z_{\text{in}}\right]}{\text{Re}(1/Z_{\text{m1}})}, \quad (20)$$

where $Z_{\text{m1}}$ can be approximated as the intrinsic impedance of medium m1, i.e., $\eta_{\text{m1}}$, and $Z_{\text{in}}$ can be obtained as [37]

$$Z_{\text{in}} = \eta_{\text{m2}}\frac{\eta_{\text{m3}} + i\eta_{\text{m2}}\tan(k_{\text{m2}}t)}{\eta_{\text{m2}} + i\eta_{\text{m3}}\tan(k_{\text{m2}}t)}. \quad (21)$$

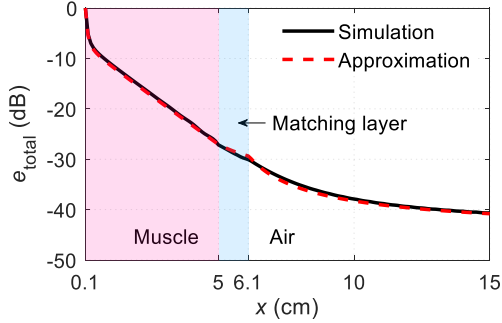

**FIGURE 9.** The total path loss $e_{\text{total}}$ as a function of the coordinate distance *x*, where the implantation depth *d* is 5 cm and a matching layer is applied.

Next, a specific case of implanted antenna is analyzed to verify the proposed analytic approximation. The implanted antenna model uses the same parameters as in Section III A with the implantation depth of 5cm. The only difference in the model is that an additional matching layer is added at the body–air interface, as shown in Fig. 7. The thickness of this quarter-wavelength matching layer is 1.1 cm, and its relative permittivity is set as 7.26. This 2-layer planar body model is numerically simulated by the full-wave simulation software FEKO. By processing the power density along the *x*-axis, we can obtain the total path loss $e_{\text{total}}$ as a function of the coordinate distance *x*, as shown in Fig. 9. Approximate results calculated from (17) are also presented and demonstrate strong agreement with simulation results. Overall, the deviation between the approximate results and the simulated results is less than 1 dB.

### C. IMPLANTED ANTENNAS IN SPHERICAL BODY MODELS

To verify the practicality of the proposed approximation method, we consider a realistic implanted antenna in a spherical body model, as shown in Fig. 10.

In this model, the host body is modeled by a spherical body phantom with the radius of $r_{\text{body}}$, which is set to 3, 5, and 7 cm in the analysis, respectively. Spherical body phantoms are commonly used to model medium-sized host bodies, especially those with significant body curvature. Thus, in modeling, $r_{\text{body}}$ represents the radius of curvature of the body interface close to the implantation location. As in the previous example, the body phantom is made of muscle.

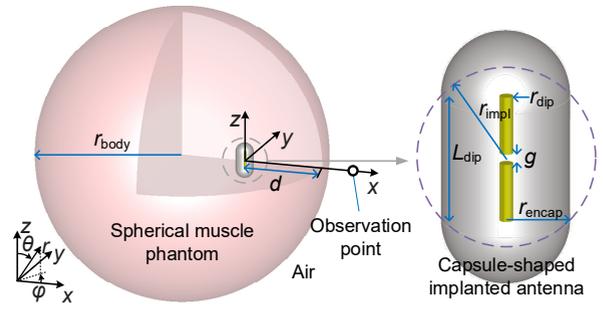

**FIGURE 10.** View of an implanted antenna in a spherical body model. The implanted antenna is modeled as a short dipole antenna surrounded by a capsule-shaped air encapsulation.

As shown in Fig. 10, a realistic 2.45-GHz capsule-shaped implanted antenna is modeled in this example. Within the capsule encapsulation, a short dipole antenna is placed in the center, consisting of two conductor wires with a radius of $r_{\text{dip}} = 0.2$ mm, a feed gap of $g = 0.2$ mm, and an overall length of $L_{\text{dip}} = 6$ mm. The capsule is filled by air, which is a cylinder (radius $r_{\text{encap}} = 3$ mm and length $L_{\text{dip}}$) terminated by two hemispherical ends with the same radius. According to the analysis of capsule-shaped implants [12], the effective radius of the implanted antenna can be approximated by the circumscribed radius of the capsule containing the conductors, i.e., $r_{\text{impl}} \approx \sqrt{L_{\text{dip}}^2/4 + r_{\text{encap}}^2} \approx 4.24$ mm. The antenna is implanted offset in the spherical body phantom with a fixed implantation depth $d = 3$ cm. The direction towards the closest body–air interface is specified as the *x*-direction. The antenna is also oriented in *z*-direction to be parallel to the closest interface.

The EM full-wave simulation solver CST Microwave Studio 2019 is used to simulate the implanted antenna model. We first study the power density of EM waves as a function of distance *x* along the positive *x*-axis. In order to eliminate the effect of feed mismatch, the simulated power density is normalized by the value at $x = r_{\text{impl}}$ as the reference value, as shown in Fig. 11.

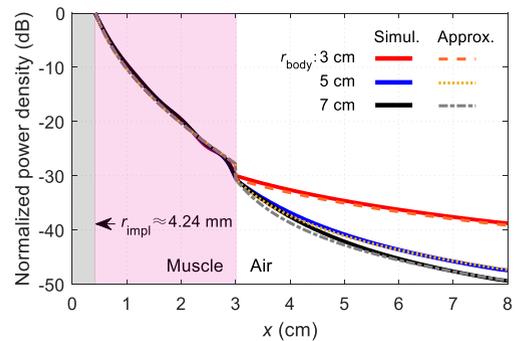

**FIGURE 11.** Normalized power density of EM waves propagating along the *x*-axis through the simulation solver and the approximate method with a fixed implantation depth *d* = 3 cm.

The power density of electromagnetic waves can also be approximately evaluated using the method proposed in this article. By transforming (1) and (2), the power density normalized by the value at $x = r_{impl}$ can be written as the function of different loss contributions, i.e.,

$$\frac{\mathrm{Re}\{S(x)\}}{\mathrm{Re}\{S(r_{impl})\}} = \begin{cases} \dfrac{r_{impl}^2}{x^2} e_{\mathrm{near\,field}} \cdot e_{\mathrm{propagation}}, & r_{impl} \leq x < d \\ \dfrac{r_{impl}^2}{x^2} e_{\mathrm{near\,field}} \cdot e_{\mathrm{propagation}} \cdot e_{\mathrm{reflection}} \cdot e_{\mathrm{refraction}}, & x \geq d \end{cases}. \quad (22)$$

Note that in calculating the power density within the body phantom ($r_{impl} \leq x < d$), $d$ in the expressions of $e_{\mathrm{near\,field}}$ and $e_{\mathrm{propagation}}$ needs to be replaced with $x$. To take into account the effects of body curvature, $e_{\mathrm{refraction}}$ needs to be calculated using (7) and (12). As shown in Fig. 11, there is a good overlap in the power densities calculated using both methods, propagating from the in-body region to free space. In particular, the attenuation process of power density in free space exhibits significant differences for different $r_{\mathrm{body}}$, i.e., the curvature radius of the body–air interface. In the case of an implant located at the center of a spherical model (i.e., $d = r_{\mathrm{body}} = 3$ cm), the power density decays the slowest in free space since it decays only by the radial spreading factor $1/(4\pi x^2)$, which is a consequence of the spherical symmetry of the considered structure.

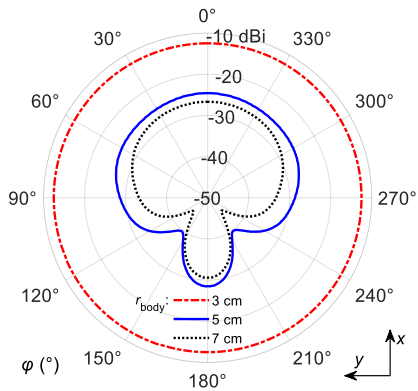

**FIGURE 12.** Simulated gain patterns of the capsule-shaped implanted antenna for spherical body phantoms with a fixed implantation depth $d = 3$ cm and different radii of curvature $r_{\mathrm{body}}$.

Furthermore, based on the proposed approximation method, we can directly evaluate the free-space path loss reaching the far-field region of the body–air interface, thereby estimating the gain of the implanted antenna. Fig. 12 shows the far-field gain patterns of the implanted antenna in the $x$-$y$ plane. The main lobe of the radiation pattern is in the $+x$-direction, which is also the direction in which the path loss is studied in Fig. 11. In estimating the gain of the implanted antenna in the $+x$-direction, i.e., $G_{+x}$, we first approximate the total path loss $e_{\mathrm{total}}$ by using the derivation of (13). Table 1 shows the approximate results of $e_{\mathrm{total}}$ for different $r_{\mathrm{body}}$. This total path loss excluding radial spreading is indeed similar in definition to the gain of the implanted antenna in the same direction. To convert $e_{\mathrm{total}}$ into the gain value in dBi, its value needs to be enlarged by 1.76 dB, which is the directivity of an electrically short dipole.

**TABLE 1.** Approximate and simulated results of the gain for a capsule-shaped antenna implanted in a spherical phantom.

| $r_{\mathrm{body}}$ | Approx. $e_{\mathrm{total}}$ | Approx. $G_{+x}$ | Simul. $G_{+x}$ |
|---|---|---|---|
| 3 cm | –14.53 dB | –12.77 dBi | –12.47 dBi |
| 5 cm | –25.40 dB | –23.64 dBi | –24.61 dBi |
| 7 cm | –27.71 dB | –25.95 dBi | –26.68 dBi |

Finally, as given in Table 1, the differences between the approximate and simulated results of $G_{+x}$ are all within 1 dB. As $r_{\mathrm{body}}$ increases, the maximum gain of the considered implanted antenna decreases and eventually approaches the maximum gain for a planar body model, i.e., –30.14 dBi.

## IV. EXPERIMENTAL APPLICATIONS

In practical applications, the proposed method can provide an effective approach to quickly estimate the wireless link efficiency between an implanted antenna and an external antenna. To illustrate its practicality, in this section, we take a wireless system prototype as an example to measure the path loss of the wireless link.

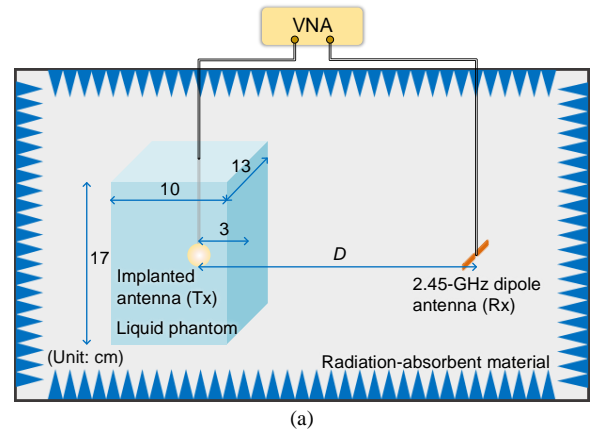

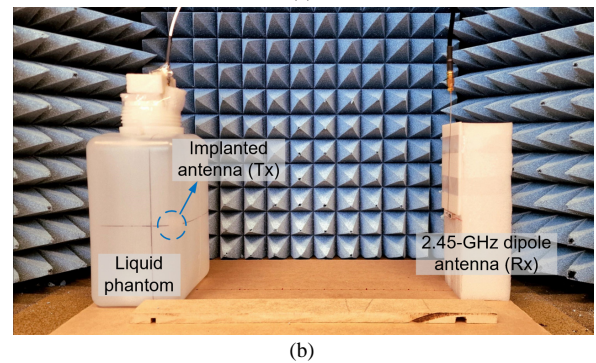

**FIGURE 13.** Measurement setup for path loss of an implanted antenna in a cubic liquid phantom. (a) Schematic diagram. (b) Photograph.

As shown in Fig. 13, this wireless system consists of an implanted antenna as the transmitter (Tx) and an external dipole antenna as the receiver (Rx), both working at 2.45

GHz. The path loss of the wireless system can be expressed by the transmission coefficient |$S_{21}$| between the two, which is measured by a vector network analyzer (VNA, 8720D HP) through coaxial cables. In order to reduce external EM interference in the measurement, radiation-absorbent materials are arranged around the antenna setup.

The implanted antenna is a 2.45-GHz meander dipole antenna encapsulated between two hemispherical ceramics ($r_{\text{impl}}$ = 7.5 mm), as illustrated in Fig. 14(a). Through printed circuit board (PCB) processing, the dipole antenna is fabricated on a 0.1-mm polyimide substrate with 18-μm thick copper metallization. To connect the balanced-fed dipole antenna to a coaxial cable, a 2.45-GHz chip balun (2450BL15B050, Johanson Technology) is introduced in the transition from the unbalanced semi-rigid cable (EZ-34, EZ Form Cable) to the balanced feeding stripline of the antenna. The hemispherical ceramics are made of aluminum oxide (Al$_2$O$_3$, $\varepsilon_r \approx 9.8$) for biocompatibility and achievable impedance matching. The specific design of the PCB and ceramic encapsulation can be seen in Fig. 14(b).

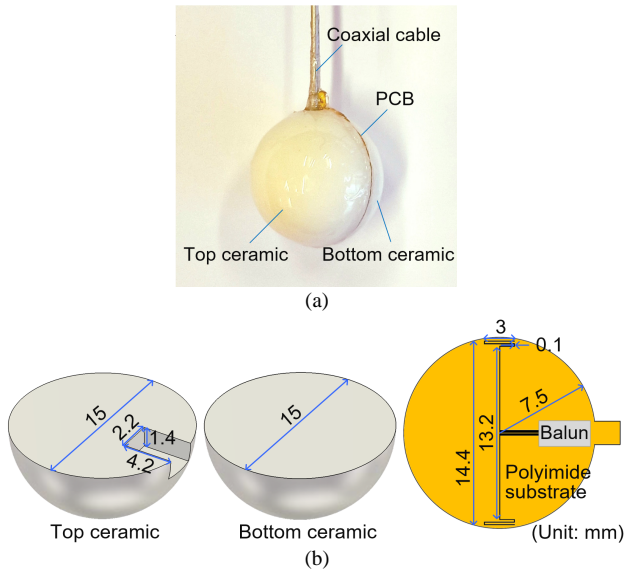

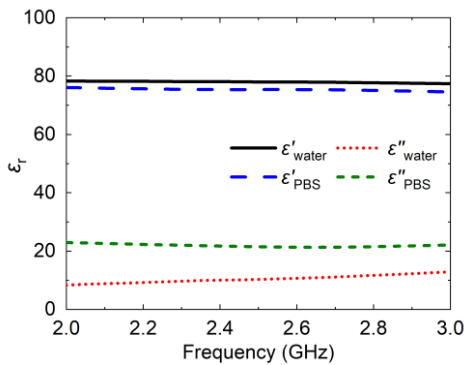

**FIGURE 14.** Ceramic-encapsulated implantable antenna used in liquid phantoms. (a) Photograph. (b) Design diagram.

**FIGURE 15.** Measured permittivity of distilled water and PBS buffer solution.

To model the host body, we used lossy liquid to make a cuboidal body phantom. The liquid is filled into a 13×10×17 cm$^3$ cuboidal plastic container. The container is made of polyethylene, and its effect on the antenna is negligible as the wall thickness is only 0.3 mm. Two lossy liquids are used in the phantom: distilled water with the relative permittivity of $\varepsilon_r \approx 78.06 - i10.10$ at 2.45 GHz and phosphate-buffered saline (PBS) (pH 7.4) buffer solution with $\varepsilon_r \approx 75.39 - i21.61$ at 2.45 GHz. The permittivity of the liquid is measured using a dielectric assessment kit (DAK-3.5, SPEAG). As depicted in Fig. 15, the measured permittivities of the two lossy liquids have similar real parts but different imaginary parts, i.e., the PBS buffer solution has more losses than distilled water.

The implanted antenna is immersed in the container and 3 cm away from the closest planar interface of the body phantom (i.e., the side wall of the container), which is the implantation depth. To maximize the gain, the implanted antenna is oriented horizontally and parallel to the planar interface. As shown in Fig. 16, the antenna's reflection coefficient |$S_{11}$| is measured by a vector network analyzer (VNA, HP 8720D), which shows that the antenna is matched well at 2.45 GHz in both lossy liquids. This is mainly attributed to the similar real parts of the permittivities of both liquids. The external dipole antenna (Rx), located in the direction from the implanted antenna to the closest interface, is a commercial antenna (SRF2W012-100, Antenova) attached to a 40×7.5×3 mm$^3$ acrylic board, with a measured gain of –2.38 dBi at 2.45 GHz. The variable distance between the implanted antenna (Tx) and the external dipole antenna (Rx) is denoted as $D$.

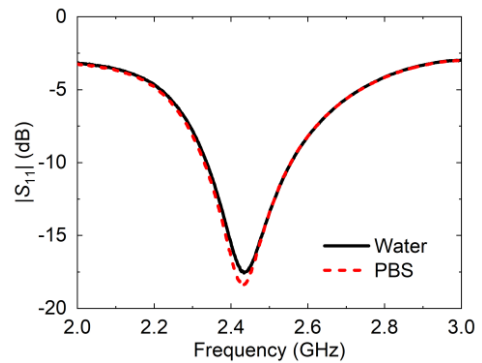

**FIGURE 16.** Measured reflection coefficients |$S_{11}$| of the implantable antenna immersed in the liquid phantom made of distilled water and PBS buffer solution, respectively.

By changing the variable distance $D$, the transmission coefficient |$S_{21}$| of the wireless system is measured for two different phantom liquids, as shown by the blue dots in Fig. 17.

According to the definition of $e_{\text{total}}$, |$S_{21}$| can be approximated by multiplying the radiated power density of the implanted antenna and the effective aperture $A_e$ of the external antenna:

$$|S_{21}|^2 = \frac{\text{Re}\{S(D)\}A_e}{4\pi r_{impl}^2 \text{Re}\{S(r_{impl})\}/g_{Tx}} = \frac{g_{Tx}A_e}{4\pi r_{impl}^2}e_{total}\frac{r_{impl}^2}{D^2}$$
$$= \frac{g_{Tx}g_{Rx}\lambda_{air}^2}{(4\pi D)^2}e_{near\,field}\cdot e_{propagation}\cdot e_{reflection}\cdot e_{refraction}, \quad (23)$$

where $g_{Tx}$ is the intrinsic gain of an electrically short dipole (typically it is equal to 1.76 dB), $A_e$ is expressed as $A_e = g_{Rx}\lambda_{air}^2/4\pi$, $g_{Rx}$ is the gain of the external antenna, and $\lambda_{air}$ is the wavelength in free space.

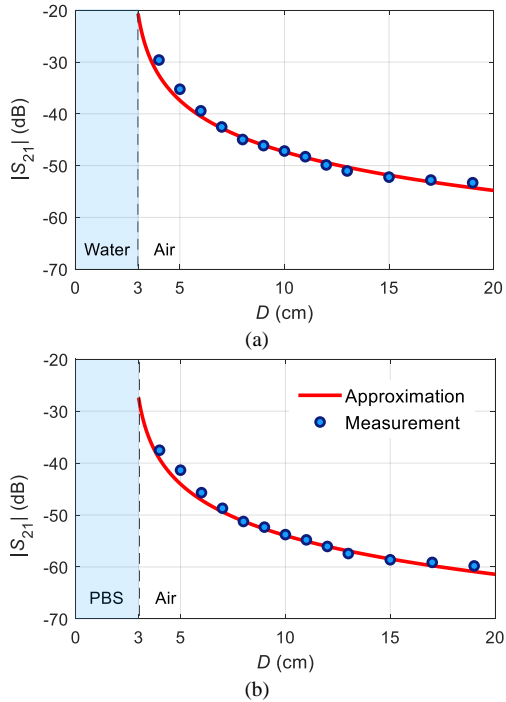

**FIGURE 17.** Measured and approximate $|S_{21}|$ as a function of the distance $D$ between the implanted antenna and the external antenna, where the liquid phantom is made of (a) distilled water and (b) PBS buffer, respectively.

As a result, the red curves in Fig. 17 illustrate the approximate $|S_{21}|$ for different phantom liquids as a function of the distance $D$. Overall, the approximate results show close agreement with the measured results, with most deviations being less than 1 dB. For observation points close to the phantom interface (i.e., $D \leq 6\,\text{cm}$), the measured results deviate from the approximate results by up to 3 dB. This is because the portion of the near field reaching the interface is not included in the approximation. As the observation point moves to the far-field region, the gain of the implanted antenna tends to a constant value. Compared with the case of the distilled water phantom, the path loss of the PBS buffer phantom is generally reduced by more than 6 dB due to the greater loss of the PBS buffer.

## V. CONCLUSION

In this work, we have proposed closed-form approximate expressions to estimate the free-space path loss for implanted antennas within large-scale host bodies. Beyond the typical radial spreading of electromagnetic waves inherent to all antennas, this free-space path loss model highlights the effect of electromagnetic scattering at the body–air interface. The proposed approximation, combined with the existing approximation of in-body path loss, allows for a quick assessment of wireless transmission efficiency from the implant to the external node.

Analysis of the free-space path loss of implanted antennas reveals the significant impact of high permittivity properties of biological tissues on losses due to wave refraction at the body–air interface. Two analytical models are investigated to analyze the refraction process of the spherical electromagnetic waves at the planar and spherical body–air interfaces, respectively. Consequently, closed-form approximate expressions are derived for estimating free-space path loss along the direction from the implant to the closest body–air interface. Notably, the free-space path loss for implanted antennas is a function of the free-space distance, the depth of the implanted antenna, and the permittivity of the lossy medium, and additionally the curvature radius of the body–air interface also affects the path loss.

The proposed approximation method is first validated on planar body models through numerical calculations of Green's functions for layered media. Then, we further investigate several realistic examples of implanted antennas through full-wave simulation analysis and prototype measurements. The results of the approximate expressions are shown to be in good agreement with simulation and measurement results, underscoring the practical utility of the method in the design of implanted antennas or implantable wireless systems.

### ACKNOWLEDGMENT
The authors are grateful for the assistance from Hannes Bartle in building the measurement setup.

### REFERENCES
[1] N. M. van Hemel and E. E. van der Wall, "8 October 1958, D Day for the implantable pacemaker," *Neth. Heart J.*, vol. 16, no. 1, pp. 1–2, Jan. 2008.

[2] D. Fitzpatrick, *Implantable Electronic Medical Devices*. New York, NY, USA: Elsevier, 2014.

[3] E. Katz, *Implantable Bioelectronics*. Weinheim, Germany: Wiley, 2014.

[4] A. N. Khan, Y.-O. Cha, H. Giddens, and Y. Hao, "Recent Advances in Organ Specific Wireless Bioelectronic Devices: Perspective on Biotelemetry and Power Transfer Using Antenna Systems," *Engineering*, vol. 11, pp. 27–41, Apr. 2022.

[5] Y. Li, N. Li, N. De Oliveira, and S. Wang, "Implantable bioelectronics toward long-term stability and sustainability," *Matter*, vol. 4, no. 4, pp. 1125–1141, Apr. 2021.

[6] K. Agarwal, R. Jegadeesan, Y.-X. Guo, and N. V. Thakor, "Wireless Power Transfer Strategies for Implantable Bioelectronics," *IEEE Rev. Biomed. Eng.*, vol. 10, pp. 136–161, 2017.

[7] S. Yoo, J. Lee, H. Joo, S. Sunwoo, S. Kim, and D. Kim, "Wireless Power Transfer and Telemetry for Implantable Bioelectronics," *Adv. Healthc. Mater.*, vol. 10, no. 17, Jun. 2021.

[8] E. Chow, M. Morris, and P. Irazoqui, "Implantable RF Medical Devices: The Benefits of High-Speed Communication and Much

Greater Communication Distances in Biomedical Applications," *IEEE Microw. Mag.*, vol. 14, no. 4, pp. 64–73, Jun. 2013.

[9] N. A. Malik, P. Sant, T. Ajmal, and M. Ur-Rehman, "Implantable Antennas for Bio-Medical Applications," *IEEE J. Electromagn. RF Microw. Med. Biol.*, vol. 5, no. 1, pp. 84–96, Mar. 2021.

[10] A. Kiourti *et al.*, "Next-Generation Healthcare: Enabling Technologies for Emerging Bioelectromagnetics Applications," *IEEE Open J. Antennas Propag.*, vol. 3, pp. 363–390, 2022.

[11] K. Guido and A. Kiourti, "Wireless Wearables and Implants: A Dosimetry Review," *Bioelectromagnetics*, vol. 41, no. 1, pp. 3–20, Dec. 2019.

[12] D. Nikolayev, W. Joseph, M. Zhadobov, R. Sauleau, and L. Martens, "Optimal Radiation of Body-Implanted Capsules," *Phys. Rev. Lett.*, vol. 122, no. 10, Mar. 2019.

[13] A. Khalifa, S. Lee, A. C. Molnar, and S. Cash, "Injectable wireless microdevices: challenges and opportunities," *Bioelectron. Med.*, vol. 7, no. 1, Dec. 2021.

[14] H. T. Friis, "A Note on a Simple Transmission Formula," *Proc. IRE*, vol. 34, no. 5, pp. 254-256, May 1946.

[15] C. A. Balanis, *Antenna Theory: Analysis and Design, 4th ed.* Hoboken, NY, USA: Wiley, 2016.

[16] H. A. Wheeler, "Fundamental Limitations of Small Antennas," *Proc. IRE*, vol. 35, no. 12, pp. 1479–1484, Dec. 1947.

[17] L. J. Chu, "Physical Limitations of Omni-Directional Antennas," *J. Appl. Phys.*, vol. 19, no. 12, pp. 1163–1175, 1948.

[18] D. M. Pozar, "Microstrip antennas," *Proc. IEEE*, vol. 80, no. 1, pp. 79-91, Jan. 1992.

[19] J. Kim and Y. Rahmat-Samii, "Implanted antennas inside a human body: Simulations, designs, and characterizations," *IEEE Trans. Microw. Theory Tech.*, vol. 52, no. 8, pp. 1934–1943, Aug. 2004.

[20] M. Gao, Z. Šipuš, I. V. Soares, S. Raman, D. Nikolayev and A. K. Skrivervik, "Analytical Model for Calculating Gain Pattern of Antennas Implanted in Large Host Bodies," *IEEE Trans. Antennas Propag.*, Early Access, 2024.

[21] A. S. Y. Poon, S. O. Driscoll, and T. H. Meng, "Optimal Frequency for Wireless Power Transmission Into Dispersive Tissue," *IEEE Trans. Antennas Propag.*, vol. 58, no. 5, pp. 1739–1750, 2010.

[22] F. Merli, B. Fuchs, J. R. Mosig, and A. K. Skrivervik, "The Effect of Insulating Layers on the Performance of Implanted Antennas," *IEEE Trans. Antennas Propag.*, vol. AP-59, no. 1, pp. 21–31, Jan. 2011.

[23] M. Manteghi and A. A. Y. Ibraheem, "On the Study of the Near-Fields of Electric and Magnetic Small Antennas in Lossy Media," *IEEE Trans. Antennas Propag.*, vol. 62, no. 12, pp. 6491–6495, Dec. 2014.

[24] A. Skrivervik, M. Bosiljevac, and Z. Šipuš, "Fundamental Limits for Implanted Antennas: Maximum Power Density Reaching Free Space," *IEEE Trans. Antennas Propag.*, vol. 67, no. 8, pp. 4978–4988, Aug. 2019.

[25] M. Gao, Z. Sipus, and A. K. Skrivervik, "Analytic Approximation of In-Body Path Loss for Implanted Antennas," *IEEE Open J. Antennas Propag.*, vol. 4, pp. 537-545, 2023.

[26] D. B. Smith, D. Miniutti, T. A. Lamahewa, and L. W. Hanlen "Propagation Models for Body-Area Networks: A Survey and New Outlook," *IEEE Ant. Propag. Mag.*, vol. 55, no. 5, pp. 97-117, Oct. 2013.

[27] D. Kurup, G. Vermeeren, E. Tanghe, W. Joseph, and L. Martens, "In-to-Out Body Antenna-Independent Path Loss Model for Multilayered Tissues and Heterogeneous Medium," *Sensors*, vol. 15, no. 1, pp. 408–421, 2015.

[28] R. Chávez-Santiago *et al.*, "Experimental Path Loss Models for In-Body Communications Within 2.36-2.5 GHz," *IEEE J. Biomed. Health Inform.*, vol. 19, no. 3, pp. 930–937, May 2015.

[29] M. Gao *et al.*, "Radiation Patterns of RF Wireless Devices Implanted in Small Animals: Unexpected Deformations Due to Body Resonance," *IEEE Trans. Biomed. Circuits Syst.*, 2023.

[30] S. Gabriel, R.W. Lau, and C. Gabriel, "The dielectric properties of biological tissues: III. Parametric models for the dielectric spectrum of tissues," *Phys. Med. Biol.*, vol. 41, 2271, 1996.

[31] A. Biggs and H. Swarm, "Radiation fields of an inclined electric dipole immersed in a semi-infinite conducting medium," *IEEE Trans. Antennas Propag.*, vol. 11, no. 3, pp. 306-310, May 1963.

[32] J. Holmes and C. Balanis, "Refraction of a uniform plane wave incident on a plane boundary between two lossy media," *IEEE Trans. Antennas Propag.*, vol. 26, no. 5, pp. 738–741, Sep. 1978.

[33] R. F. Harrington, *Time-Harmonic Electromagnetic Fields*. New York, NY, USA: McGraw-Hill, 1961.

[34] J. A. Kong, *Theory of electromagnetic waves*. New York, Wiley-Interscience, 1975.

[35] J. R. Mosig, "Integral Equation Technique," *Numerical Techniques for Microwave and Millimeter Wave Passive Structures*, T. Itoh, Ed. New York: Wiley, 1989.

[36] K. A. Michalski and J. R. Mosig, "Multilayered media Green's functions in integral equation formulations," *IEEE Trans. Antennas Propag.*, vol. 45, no. 3, pp. 508-519, March 1997.

[37] D. M. Pozar, *Microwave Engineering*, 4th ed. Hoboken, NJ, USA: Wiley, 2012.

[38] C. A. Balanis, *Advanced Engineering Electromagnetics*, 2nd ed. Hoboken, NJ, USA: Wiley, 2012.

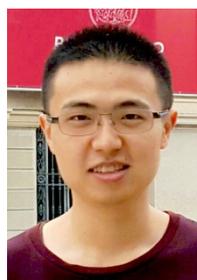

**MINGXIANG GAO** received the B.Sc. degree and the first M.Sc. degree in electrical engineering from Xi'an Jiaotong University, Xi'an, China, in 2016 and 2019, respectively, the second M.Sc. degree (summa cum laude) in electrical engineering from the Politecnico di Milano, Milan, Italy, in 2019, and the Ph.D. degree in electrical engineering from the Ecole Polytechnique Fédérale de Lausanne, Lausanne, Switzerland, in 2024. He is currently a Postdoctoral Researcher with the Foundation for Research on Information Technologies in Society, Zürich, Switzerland. His research interests include the theory and design of implantable antennas, bioelectromagnetics, and wireless bioelectronics such as neural implants.

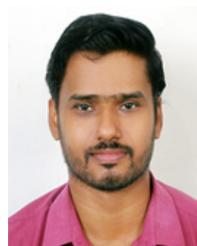

**SUJITH RAMAN** (Senior Member, IEEE) received the master's degree in electronic and the Ph.D. degree in microwave electronics from Cochin University of Science and Technology (CUSAT), India, in 2007, and 2012, respectively. Presently, he is an Assistant professor with Radio Systems group, University of Twente, Netherlands. Prior to this he was a Marie curie fellow under the MSCA Individual Fellowship at Microwaves and Antennas group, EPFL, Switzerland. He was working as UGC-Assistant professor at Bharathiar University, India. from March 2016 under the UGC-Faculty recharge program (FRP). He was with solid-state electronics department, Angstrom Laboratory, Uppsala University, Sweden as a post-doctoral researcher and also with WiSAR lab, Ireland, as a postdoctoral researcher. He is the author or coauthor of more than 80 publications including journals, conferences and book. His research interest includes antennas, biodegradable microwave devices, bio electromagnetics and microwave material characterization. Dr. Raman is a senior member of the IEEE and URSI from 2019 and 2021 onwards. He received the URSI Young Scientist Award during URSI General Assembly and Scientific Symposium 2011, Istanbul, Turkey and URSI –RCRS -2017 at Tirupati, India.

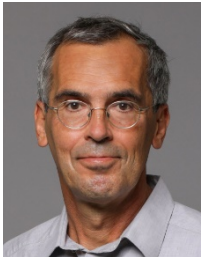

**ZVONIMIR SIPUS** (Senior Member, IEEE) received the B.Sc. and M.Sc. degrees in electrical engineering from the University of Zagreb, Zagreb, in 1988 and 1991, respectively, and the Ph.D. degree in electrical engineering from the Chalmers University of Technology, Gothenburg, Sweden, in 1997. From 1988 to 1994, he was with Rudjer Boskovic Institute, Zagreb, as a Research Assistant, where he was involved in the development of detectors for explosive gasses. In 1994, he joined the Antenna Group, Chalmers University of Technology, where he was involved in research projects concerning conformal antennas and soft and hard surfaces. In 1997, he joined the Faculty of Electrical Engineering and Computing, University of Zagreb, where he is currently a Professor. From 1999 to 2005, he was also an Adjunct Researcher with the Department of Electromagnetics, Chalmers University of Technology. Since 2006, he has been involved in teaching with the European School of Antennas. From 2008 to 2012 and from 2014 to 2018, he was the Head of the Department of Wireless Communications. His current research interests include the analysis and design of electro-magnetic structures with application to antennas, microwaves, and optical communication and sensor systems.

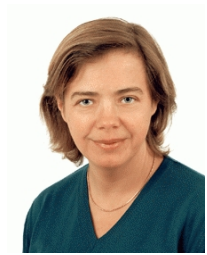

**ANJA K. SKRIVERVIK** received the master's degree in electrical engineering and the Ph.D. degree from the Ecole Polytechnique Fédérale de Lausanne (EPFL), Lausanne, Switzerland, in 1986 and 1992, respectively. She was an Invited Research Fellow with the University of Rennes, Rennes, France, followed by two years in the industry. In 1996, she rejoined EPFL as an Assistant Professor, where she is currently a Full Professor and also the Head of the Microwaves and Antennas Group. She was the Director of the EE section from 1996 to 2000, and is currently the Director of the EE Doctoral School with EPFL. Her teaching activities include courses on microwaves and antennas and courses at Bachelor, Master, and Ph.D. levels. She has authored or coauthored more than 200 peer-reviewed scientific publications. Her current research interests include electrically small antennas, antennas in biological media, multifrequency and ultrawideband antennas, and numerical techniques for electromagnetics. She was the recipient of the Latsis Award. She is frequently requested to review research programs and centres in Europe. She was the Chairperson of the Swiss URSI until 2012. She is very active in European collaboration and European Projects. Since 2017, she has been a member of the Board of Directors of the European Association on Antennas and Propagation and is a Board Member of the European School on Antennas.